\documentclass[twocolumn,showpacs,floats,floatfix,bm,superscriptaddress,aps,pra]{revtex4}
\usepackage{amsfonts}
\usepackage{amssymb}
\usepackage{amsmath}
\usepackage{graphicx}
\usepackage{bm}

\begin{document}

\author{P. A. Ivanov}
\affiliation{Department of Physics, Sofia University, James Bourchier 5 blvd, 1164 Sofia,
Bulgaria}
\author{B. T. Torosov}
\affiliation{Department of Physics, Sofia University, James Bourchier 5 blvd, 1164 Sofia,
Bulgaria}
\author{N. V. Vitanov}
\affiliation{Department of Physics, Sofia University, James Bourchier 5 blvd, 1164 Sofia,
Bulgaria}
\affiliation{Institute of Solid State Physics, Bulgarian Academy of Sciences,
Tsarigradsko chauss\'{e}e 72, 1784 Sofia, Bulgaria}
\title{Navigation between quantum states by quantum mirrors}
\date{\today }

\begin{abstract}
We introduce a technique which allows one to connect \emph{any} two
arbitrary (pure or mixed) superposition states of an $N$-state quantum
system. The proposed solution to this inverse quantum mechanical problem is
analytical, exact and very compact. The technique uses standard and
generalized quantum Householder reflections (QHR) [Ivanov \emph{et al},
Phys. Rev. A \textbf{74}, 022323 (2006)], which require external pulses of
precise areas and frequencies. We show that any two pure states can be
linked by two standard QHRs, or by only one generalized QHR. The transfer
between any two mixed states with the same dynamic invariants (e.g., the
same eigenvalues of the density matrix $\mathbf{\rho }$) requires in general 
$N$ QHRs. Moreover, we propose recipes for synthesis of arbitrary
preselected mixed states, starting in a single basis state and using a
combination of QHRs and incoherent processes (pure dephasing or spontaneous
emission).
\end{abstract}

\maketitle

\section{Introduction}

Quantum state engineering in atoms and molecules traditionally uses three
basic techniques for transfer of population, complete or partial, from one
bound energy state to another, single or superposition state: resonant
pulses of precise areas (e.g. $\pi $\ pulses in a two-state system or
generalized $\pi $ pulses for multiple states) \cite{Shore}, adiabatic
passage using one or more level crossings \cite{ARPC}, or stimulated Raman
adiabatic passage (STIRAP) and its extensions \cite{STIRAP}. All these
techniques require the system to be initially in a single energy state; such
a state can be easily prepared experimentally, e.g. by optical pumping. Some
of these techniques are \textquotedblleft tuned\textquotedblright\ to a
specific initial condition: for example, STIRAP requires a counterintuitive
pulse sequence to transfer population from state 1 to 3 in a 1-2-3 linkage,
but it is largely irrelevant if the system starts in states 2 or 3 (with
some exceptions for state 3) \cite{STIRAP}. In other words, STIRAP is (very)
useful in producing only one column (the first) of the unitary propagator.
Similar conclusions apply, to a large extent, also to the other two
techniques using pulse areas and level crossings.

These traditional techniques resolve only a small (although important) part
of the general problem of quantum state engineering: given the initial and
final states of an $N$-state system, find a physical set of operations that
connect them. This problem requires the construction of the \emph{entire
propagator}, not just a single column or row.

In this paper we introduce a technique for full quantum state engineering,
which produces in a systematic manner a propagator that can connect any two
preselected superposition states of an $N$-state quantum system,
representing a \emph{qunit} in quantum information \cite{QI}. The two states
can be pure as well as mixed, and the latter may have the same or different
sets of dynamic invariants (constants of motion). The solution consists of
two steps: first, find a propagator that connects the two states, and
second, find a physical realization of this propagator.

The \emph{first} part is the mathematical solution of this inverse problem
in quantum mechanics, and the solution is different for three types of
problems: (i) pure-to-pure states; (ii) mixed-to-mixed states with the same
invariants; (iii) mixed-to-mixed states with different invariants. The case
(iii), for instance, contains the important problem of engineering an
arbitrary presected mixed state and we pay special attention to it. In this
latter respect our \emph{exact analytic} results are alternative to the
(approximate) numeric optimization procedure proposed by Karpati \emph{et al}
\cite{Karpati}; moreover, our approach allows one to engineer \emph{any}
preselected mixed state, whereas the method of Karpati \emph{et al} \cite%
{Karpati} can only produce a class of mixed states.

The \emph{second} part of the solution is the physical realization of the
respective propagator. For this we use the recently introduced physical
implementation of the quantum Householder reflection (QHR) \cite%
{Kyoseva,Ivanov} and we show that QHR is a very powerful tool for quantum
state engineering. Remarkably, in case (i) only a single QHR is needed to
connect two pure states. In case (ii), a general U($N$) propagator is
necessary in the general case, which requires $N$ QHRs. In case (iii), some
sort of incoherent process is required in order to equalize the different
dynamic invariants of the initial and final mixed states, and the remaining
coherent U($N$) part is realized by QHRs. We describe the use of two such
incoherent processes: pure dephasing and spontaneous emission.

\begin{figure}[tbp]
\includegraphics[width=60mm]{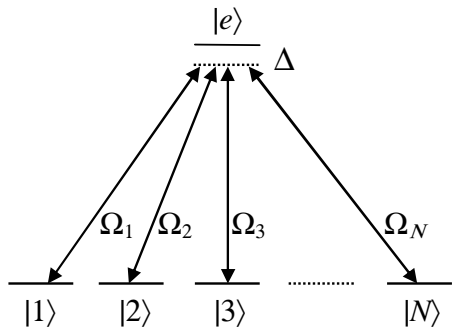}
\caption{Physical realization of the quantum Householder reflection: $N$
degenerate (in RWA sense) ground states, forming the \emph{qunit},
coherently coupled via a common excited state by pulsed external fields of
the same time dependence and the same detuning, but possibly different
amplitudes and phases. }
\label{Fig-Npod}
\end{figure}

The Householder reflection \cite{Householder} is a powerful and numerically
very robust unitary transformation, which has many applications in classical
data analysis, e.g., in solving systems of linear algebraic equations,
finding eigenvalues of high-dimensional matrices, least-square optimization,
QR decomposition, etc. \cite{Householder applications}. In its quantum
mechanical implementation \cite{Kyoseva,Ivanov} it consists of a single
interaction step involving $N$\ simultaneous pulsed fields of precise areas
and detunings in an $N$-pod linkage pattern, wherein the $N$ states of our
system are coupled to each other via an ancillary excited state, as
displayed in Fig. \ref{Fig-Npod}. We use two types of QHRs: standard and
generalized; the latter involves an additional phase factor. The standard
QHR can operate on or off resonance, whereas the generalized QHR requires
specific detunings. \emph{Any} unitary matrix can be decomposed into (and
therefore, synthesized by) $N-1$\ standard QHRs and a phase gate, or into $N$%
\ generalized QHRs, without a phase gate; hence only $N$ physical operations
are needed, which allows one to greatly reduce the number of physical steps,
from $O(N^{2})$\ in existing U(2) realizations \cite{qudits-SU(2)} to only $%
O(N)$\ with QHRs.

%
%
%
%
%
%
%
%
%

This paper is organized as follows. In Sec. \ref{Sec-QHR} we review the
standard and generalized QHR gates and their physical implementations. In
Sec. \ref{Sec-pure} we show how two pure states can be connected by means of
standard and generalized QHRs. In Sec. \ref{Sec-mixed} we construct the
propagator connecting two arbitrary mixed states with the same dynamic
invariants. Engineering of an arbitrary preselected mixed qunit state is
presented in Sec. \ref{Sec-engineering}. The conclusions are summarized in
Sec. \ref{Sec-conclusions}.

\section{The tool: Quantum Householder Reflection \label{Sec-QHR}}

\subsection{Definition}

The \emph{standard} QHR is defined as%
\begin{equation}
\mathbf{M}(v)=\mathbf{I}-2\left\vert v\right\rangle \left\langle
v\right\vert ,  \label{SQHR}
\end{equation}%
where $\mathbf{I}$ is the identity operator and $\left\vert v\right\rangle $
is an $N$-dimensional normalized complex column-vector. The QHR (\ref{SQHR})
is both hermitian and unitary, $\mathbf{M}=\mathbf{M}^{^{\dagger }}=\mathbf{M%
}^{-1}$, which means that $\mathbf{M}$ is involutary, $\mathbf{M}^{2}=%
\mathbf{I}$. In addition, $\det \mathbf{M}=-1$. For real $\left\vert
v\right\rangle $ the Householder transformation (\ref{SQHR}) has a simple
geometric interpretation: reflection with respect to an $(N-1)$-dimensional
plane with a normal vector $\left\vert v\right\rangle $. In general, the
vector $\left\vert v\right\rangle $ is complex and it is characterized by $%
2N-2$ real parameters (with the normalization condition and the unimportant
global phase accounted for).

The \emph{generalized} QHR is defined as%
\begin{equation}
\mathbf{M}(v;\varphi )=\mathbf{I}+\left( e^{i\varphi }-1\right) \left\vert
v\right\rangle \left\langle v\right\vert ,  \label{GQHR}
\end{equation}%
where $\varphi $ is an arbitrary phase. The standard QHR (\ref{SQHR}) is a
special case of the generalized QHR (\ref{GQHR}) for $\varphi =\pi $: $%
\mathbf{M}(v;\pi )\equiv \mathbf{M}(v)$. The generalized QHR is unitary, $%
\mathbf{M}(v;\varphi )^{-1}=\mathbf{M}(v;\varphi )^{\dagger }=\mathbf{M}%
(v;-\varphi )$, and its determinant is $\det \mathbf{M}=e^{i\varphi }$.

\subsection{Physical implementation}

We have shown recently \cite{Kyoseva,Ivanov} that the standard and
generalized QHR operators can be realized physically in a coherently coupled 
$N$-pod system shown in Fig. \ref{Fig-Npod}. The $N$ degenerate [in the
rotating-wave approximation (RWA) sense \cite{Shore}] ground states $%
\left\vert n\right\rangle $ ($n=1,2,\ldots ,N$), which represent the \emph{%
qunit}, are coupled coherently by $N$ external fields to an ancillary,
excited state $\left\vert e\right\rangle \equiv \left\vert N+1\right\rangle $
\cite{Kyoseva}. The excited state $\left\vert e\right\rangle $ can generally
be off resonance by a detuning $\Delta \left( t\right) $ \cite{Kyoseva},
which must be the same for all fields. The Rabi frequencies $\Omega
_{1}(t),\ldots ,\Omega _{N}(t)$ of the couplings between the ground states
and the excited state have the same pulse-shaped time dependence $f\left(
t\right) $, but possibly different phases $\beta _{n}$ and amplitudes $\chi
_{n}$, 
\begin{equation}
\Omega _{n}(t)=\chi _{n}f\left( t\right) e^{i\beta _{n}}\text{\quad }%
(n=1,2,\ldots ,N).  \label{Omega}
\end{equation}%
The\ qunit+ancilla RWA Hamiltonian reads%
\begin{equation}
\mathbf{H}(t)=\frac{\hbar }{2}\left[ 
\begin{array}{ccccc}
0 & 0 & \cdots  & 0 & \Omega _{1}\left( t\right)  \\ 
0 & 0 & \cdots  & 0 & \Omega _{2}\left( t\right)  \\ 
\vdots  & \vdots  & \ddots  & \vdots  & \vdots  \\ 
0 & 0 & \cdots  & 0 & \Omega _{N}\left( t\right)  \\ 
\Omega _{1}^{\ast }\left( t\right)  & \Omega _{2}^{\ast }\left( t\right)  & 
\cdots  & \Omega _{N}^{\ast }\left( t\right)  & 2\Delta \left( t\right) 
\end{array}%
\right] ,  \label{Hamiltonian}
\end{equation}%
The exact solution to the Schr\"{o}dinger equation for the propagator $%
\mathbf{U}(t)$,%
\begin{equation}
i\hbar \frac{d}{dt}\mathbf{U}(t)=\mathbf{H}(t)\mathbf{U}(t).  \label{SEq}
\end{equation}
can be found in \cite{Kyoseva}. 

The \emph{standard QHR }$\mathbf{M}(v)$ is realized on exact resonance ($%
\Delta =0$), for any pulse shape $f\left( t\right) $, and for
root-mean-square (rms) pulse area%
\begin{equation}
A=2\left( 2k+1\right) \pi \quad \left( k=0,1,2,\ldots \right) ,
\label{A-resonance}
\end{equation}%
where%
\begin{equation}
A=\int_{-\infty }^{\infty }\Omega (t)dt,  \label{rms area}
\end{equation}%
with $\Omega (t)=\left[ \sum_{n=1}^{N}\left\vert \Omega _{n}(t)\right\vert
^{2}\right] ^{1/2}$. Then the transition probabilities to the ancilla state
vanish and the propagator within the qunit space is given exactly by the
standard QHR $\mathbf{M}(v)$ (\ref{SQHR}). The components of the $N$%
-dimensional normalized complex vector $\left\vert v\right\rangle $ are the
Rabi frequencies, with the accompanying phases \cite{Ivanov},%
\begin{equation}
\left\vert v\right\rangle =\frac{1}{\chi }\left[ \chi _{1}e^{i\beta
_{1}},\chi _{2}e^{i\beta _{2}},\ldots ,\chi _{N}e^{i\beta _{N}}\right] ^{T},
\label{V}
\end{equation}%
where $\chi =\left( \sum_{n=1}^{N}\chi _{n}^{2}\right) ^{1/2}$. Hence the
qunit propagator represents a physical realization of the standard QHR\emph{%
\ }in a single interaction step. Any QHR vector (\ref{V}) can be produced on
demand by appropriately selecting the peak couplings $\chi _{n}$ and the
phases $\beta _{n}$ of the external fields.

The \emph{generalized QHR} $\mathbf{M}(v;\varphi )$ can be realized in the
same $N$-pod system, but for specific detunings off resonance. Again the
transition probabilities to the ancilla state must vanish; the corresponding
rms pulse areas (\ref{rms area}) in general depend on the pulse shape and
differ from the resonance values (\ref{A-resonance}). The propagator within
the qunit space is the generalized QHR (\ref{GQHR}), wherein the phase $%
\varphi $ depends on the interaction parameters. Although the parameters
(i.e. the rms area and the detuning) of any needed generalized QHR can be
found numerically for essentially any pulse shape, it is very convenient to
use a hyperbolic-secant pulse shape, for which there is a simple exact
analytic solution: the Rosen-Zener model \cite{RZ}, 
\begin{subequations}
\label{parameters RZ}
\begin{eqnarray}
f\left( t\right)  &=&\text{sech}\left( t/T\right) ,  \label{pshape} \\
\Delta \left( t\right)  &=&\Delta _{0}.  \label{detuning}
\end{eqnarray}%
For this pulse shape, the rms area (\ref{rms area}) is $A=\pi \chi T$. A
generalized QHR transformation $\mathbf{M}(v;\varphi )$ (\ref{GQHR}) is
realized when the interactions satisfy again Eq. (\ref{V}), and the pulse
area and the detuning obey \cite{Kyoseva,Ivanov} 
\end{subequations}
\begin{subequations}
\begin{eqnarray}
A &=&2\pi l\quad \left( l=1,2,\ldots \right) ,  \label{RZ area} \\
\varphi  &=&2\arg \prod_{k=0}^{l-1}\left[ \Delta _{0}T+i\left( 2k+1\right) %
\right] .  \label{RZ detuning}
\end{eqnarray}%
For any given $\varphi $, there are $l$ values of $\Delta _{0}$, which
satisfy Eq. (\ref{RZ detuning}) \cite{Kyoseva}. This is also the case for $%
\varphi =\pi $, i.e. for the standard QHR, for which one of the solutions is 
$\Delta _{0}=0$. Hence the standard QHR $\mathbf{M}(v)$ can be realized both
on and off resonance, whereas the generalized QHR $\mathbf{M}(v;\varphi )$
can only be realized for nonzero $\Delta _{0}$. The advantage of tuning off
resonance is the lower transient population in the ancilla excited state,
which would reduce the population losses if the lifetime of this state is
short compared to the interaction duration.

This implementation is particularly suited for a qutrit ($N=3$) formed of
the magnetic sublevels of an atomic level with angular momentum $J=1$; then
the ancilla state should be a $J=0$ state. The three pulsed fields can be
delivered from the same laser by using beam splitters and polarizers, which
would authomatically ensure that all of them have the same detuning and
pulse shape. Moreover, with femtosecond pulses it would be possible to use
pulse shapers \cite{femto}, which can easily deliver pulses with the desired
areas. Of course, the use of femtosecond pulses offers another advantage:
decoherence is irrelevant on such time scales.

\subsection{Householder decomposition of a U($N$) propagator}

The standard QHR $\mathbf{M}(v)$ and the generalized QHR $\mathbf{M}%
(v;\varphi )$ can be used for U($N$) decomposition \cite{Ivanov}. Any $N$%
-dimensional unitary matrix $\mathbf{U}$ ($\mathbf{U}^{-1}=\mathbf{U}%
^{\dagger }$) can be expressed as a product of $N-1$ standard QHRs $\mathbf{M%
}\left( v_{n}\right) $ $(n=1,2,\ldots N-1)$ and a phase gate, 
\end{subequations}
\begin{equation}
\Phi \left( \phi _{1},\ldots ,\phi _{N}\right) =\sum_{n=1}^{N}e^{i\phi
_{n}}\left\vert n\right\rangle \left\langle n\right\vert =\text{diag}\left\{
e^{i\phi _{1}},\ldots ,e^{i\phi _{N}}\right\} ,  \label{phase gate}
\end{equation}%
as%
\begin{equation}
\mathbf{U=M}\left( v_{1}\right) \mathbf{M}\left( v_{2}\right) \ldots \mathbf{%
M}\left( v_{N-1}\right) \Phi \left( \phi _{1},\phi _{2},\ldots ,\phi
_{N}\right) ,  \label{standard}
\end{equation}%
or as a product of $N$ generalized QHRs, 
\begin{equation}
\mathbf{U=M}\left( v_{1};\varphi _{1}\right) \mathbf{M}\left( v_{2};\varphi
_{2}\right) \ldots \mathbf{M}\left( v_{N};\varphi _{N}\right) .
\label{generalized}
\end{equation}

\section{Transition between pure states\label{Sec-pure}}

The designed recipe for constructing a general U($N$) transformation makes
it possible to solve the important quantum mechanical problem of transfering
an $N$-state quantum system from one arbitrary preselected initial
superposition state to another such state, i.e. the inverse problem of
quantum-state engineering. The cases of pure and mixed states require
separate analyses.

\subsection{Transition by standard QHR}

\subsubsection{General case}

A pure qunit state\emph{\ }is described by a state vector $\left\vert \Psi
\right\rangle =\sum_{n=1}^{N}c_{n}\left\vert n\right\rangle $, where the
vectors $\left\vert n\right\rangle $ represent the qunit basis states, and $%
c_{n}$ is the complex-valued probability amplitude of state $\left\vert
n\right\rangle $. Given the preselected initial state $\left\vert \Psi
_{i}\right\rangle $ and the final state $\left\vert \Psi _{f}\right\rangle $
of the qunit, we wish to find a propagator $\mathbf{U}$, such that%
\begin{equation}
\left\vert \Psi _{f}\right\rangle =\mathbf{U}\left\vert \Psi
_{i}\right\rangle .  \label{cf=Uci}
\end{equation}

We shall show that one of the possible solutions of Eq. (\ref{cf=Uci}) reads%
\begin{equation}
\mathbf{U}=\mathbf{M}(v_{f})\mathbf{DM}(v_{i}),  \label{Us}
\end{equation}%
where $\mathbf{M}(v_{i})$ and $\mathbf{M}(v_{f})$ are standard QHRs. Here $%
\mathbf{D}$ is an $N$-dimensional unitary matrix, which, when acting upon a
single qunit basis state $\left\vert n\right\rangle $, only shifts its phase,%
\begin{equation}
\mathbf{D}\left\vert n\right\rangle =e^{i\phi _{n}}\left\vert n\right\rangle
.  \label{D}
\end{equation}%
For example, $\mathbf{D}$ can be an arbitrary $N$-dimensional phase gate (%
\ref{phase gate}). Alternatively, $\mathbf{D}$ can be an arbitrary $N$%
-dimensional generalized QHR $\mathbf{M}\left( v;\varphi \right) $ with
vector $\left\vert v\right\rangle $ orthogonal to the qunit state $%
\left\vert n\right\rangle $, $\langle v\left\vert n\right\rangle =0$.
Finally $\mathbf{D}$ can be the identity $\mathbf{D}=\mathbf{I}$.

In order to prove Eq. (\ref{Us}) we first define the vector%
\begin{equation}
\left\vert v_{\alpha n}\right\rangle =\frac{\left\vert \Psi _{\alpha
}\right\rangle -e^{i\varphi _{\alpha n}}\left\vert n\right\rangle }{\sqrt{2%
\left[ 1-\text{Re}\left( \langle \Psi _{\alpha }\left\vert n\right\rangle
e^{i\varphi _{\alpha n}}\right) \right] }},  \label{vi}
\end{equation}%
where $\left\vert n\right\rangle $ is an arbitrarily chosen basis qunit
state, $\varphi _{\alpha n}=\arg [\Psi _{\alpha }]_{n}$ and $\alpha =i,f$.
The QHR $\mathbf{M}(v_{in})$ acting upon $\left\vert \Psi _{i}\right\rangle $
reflects it onto the single qunit state $\left\vert n\right\rangle $, 
\begin{equation}
\mathbf{M}(v_{in})\left\vert \Psi _{i}\right\rangle =e^{i\varphi
_{in}}\left\vert n\right\rangle .  \label{Mi}
\end{equation}%
The action of $\mathbf{D}$ upon $\left\vert n\right\rangle $ only shifts its
phase, see Eq. (\ref{D}). The action of $\mathbf{M}(v_{fn})$ upon $%
\left\vert n\right\rangle $ reflects this vector onto the final state, 
\begin{equation}
\mathbf{M}(v_{fn})\left\vert n\right\rangle =e^{-i\varphi _{fn}}\left\vert
\Psi _{f}\right\rangle .  \label{Mfe1}
\end{equation}%
Equations (\ref{Mi}), (\ref{D}) and (\ref{Mfe1}) imply that%
\begin{equation}
\mathbf{M}(v_{fn})\mathbf{DM}(v_{in})\left\vert \Psi _{i}\right\rangle
=e^{i\left( \varphi _{in}-\varphi _{fn}+\phi _{n}\right) }\left\vert \Psi
_{f}\right\rangle ,  \label{solution}
\end{equation}%
which, up to an unimportant phase, proves Eq. (\ref{Us}).

The arbitrariness in the choice of the unitary matrix $\mathbf{D}$, and the
intermediate basis state $\left\vert n\right\rangle $, means that the
solution (\ref{Us}) for $\mathbf{U}$ is \emph{not unique}. However, what is
important is that it always exists. In fact the availability of multiple
solutions offers some flexibility for a physical implementation. In
particular we can always choose $\mathbf{D}=\mathbf{I}$; then the physical
realization of the propagator $\mathbf{U}$ requires \emph{only two} \emph{%
standard} \emph{QHRs}.

\subsubsection{Special cases}

In several important special cases \emph{only a single standard QHR} is
needed for pure-to-pure transition.

1. If the qunit is in a \emph{single initial} basis state $\left\vert \Psi
_{i}\right\rangle =\left\vert n\right\rangle $ then, as follows from Eq. (%
\ref{Mfe1}), only one standard QHR $\mathbf{M}(v_{fn})$ is sufficient to
transfer it into an arbitrary superposition state $\left\vert \Psi
_{f}\right\rangle $, with $\left\vert v_{fn}\right\rangle $ given by Eq. (%
\ref{vi}).

2. Likewise, an arbitrary inital superposition state $\left\vert \Psi
_{i}\right\rangle $ can be linked to any \emph{single final} state $%
\left\vert \Psi _{f}\right\rangle =\left\vert n\right\rangle $ by only one
standard QHR $\mathbf{M}(v_{in})$, with $\left\vert v_{in}\right\rangle $
given by Eq. (\ref{vi}).

3. If $\left\vert \Psi _{i}\right\rangle $ and $\left\vert \Psi
_{f}\right\rangle $ are \emph{orthogonal} ($\langle \Psi _{f}\left\vert \Psi
_{i}\right\rangle =0$), then again only a single standard QHR $\mathbf{M}(v)$%
, with $\left\vert v\right\rangle =\frac{1}{\sqrt{2}}\left( \left\vert \Psi
_{f}\right\rangle -\left\vert \Psi _{i}\right\rangle \right) $, is
sufficient to connect them.

4. If $\left\vert \Psi _{i}\right\rangle $ and $\left\vert \Psi
_{f}\right\rangle $ are superpositional states with \emph{real }%
coefficients, then again a single standard QHR $\mathbf{M}(v)$, with $%
\left\vert v\right\rangle =\left( \left\vert \Psi _{f}\right\rangle
-\left\vert \Psi _{i}\right\rangle \right) /\sqrt{2\left( 1-\langle \Psi
_{f}\left\vert \Psi _{i}\right\rangle \right) }$, is sufficient to link them.

\subsection{Transition by generalized QHR}

Generalized QHR is ideally suited for a pure-to-pure transition because, as
it is easily seen, \emph{only one} \emph{generalized} \emph{QHR} is
sufficient to reflect state $\left\vert \Psi _{i}\right\rangle $ onto $%
\left\vert \Psi _{f}\right\rangle $,%
\begin{equation}
\mathbf{U}=\mathbf{M}(v;\varphi ),  \label{gQHR}
\end{equation}%
where 
\begin{subequations}
\begin{eqnarray}
\left\vert v\right\rangle  &=&\frac{\left\vert \Psi _{f}\right\rangle
-\left\vert \Psi _{i}\right\rangle }{\sqrt{2\left( 1-\text{Re}\left\langle
\Psi _{f}|\Psi _{i}\right\rangle \right) }}, \\
\varphi  &=&2\arg \left( 1-\langle \Psi _{f}\left\vert \Psi
_{i}\right\rangle \right) +\pi .
\end{eqnarray}%
In comparison with (\ref{Us}) the solution (\ref{gQHR}) is unique; there is
no arbitrariness in the choice of the QHR vector $\left\vert v\right\rangle $
(up to an unimportant global phase) and the phase $\varphi $.

\subsection{Examples}

%
%
%
%
%
%
%
%
%
\begin{figure}[tbp]
\includegraphics[height=90mm]{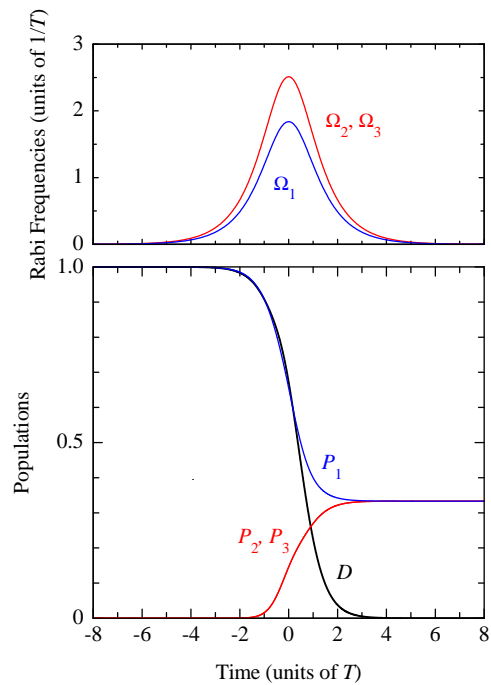}
\caption{(color online) Time evolution of the pulsed fields (top) and the
populations (bottom) for the transition (\protect\ref{1-3}) in a qutrit. We
have assumed sech pulses and rms pulse area $A=4\protect\pi $ ($\protect\chi %
T=4$). The individual couplings $\protect\chi _{n}$ $(n=1,2,3)$ are given by
the components of the QHR vector (\protect\ref{vector1}), each multiplied by 
$\protect\chi $. The detuning is $\Delta _{0}T$ $=1.732$ (which gives $%
\protect\varphi =\protect\pi $). The thick curve is the state mismatch (%
\protect\ref{mismatch}).}
\label{Fig 1-3}
\end{figure}

We consider a \emph{qutrit} ($N=3$), for which the QHR implementation is
particularly suitable. As a first example, the transition from a single
state to a superposition state, 
\end{subequations}
\begin{equation}
\left\vert \Psi _{i}\right\rangle =\left\vert 1\right\rangle \longrightarrow 
\frac{\left\vert 1\right\rangle +\left\vert 2\right\rangle +\left\vert
3\right\rangle }{\sqrt{3}}=\left\vert \Psi _{f}\right\rangle ,  \label{1-3}
\end{equation}%
is performed by a \emph{single QHR }$\mathbf{M}(v)$, with 
\begin{equation}
\left\vert v\right\rangle =\frac{1}{2}\sqrt{1+\frac{1}{\sqrt{3}}}\left[ 
\sqrt{3}-1,-1,-1\right] ^{T}.  \label{vector1}
\end{equation}%
Figure \ref{Fig 1-3} shows the corresponding time evolutions of the
populations and the state mismatch $D$. The latter is defined as the
distance between the qutrit state vector $\left\vert \Psi (t)\right\rangle $
and the target state $\left\vert \Psi _{f}\right\rangle $,%
\begin{equation}
D(t)=\frac{\sum_{mn}\left\vert \rho _{mn}(t)-\rho _{mn}^{f}\right\vert }{%
\sum_{mn}\left\vert \rho _{mn}^{i}-\rho _{mn}^{f}\right\vert },
\label{mismatch}
\end{equation}%
where $\rho _{mn}$ are the elements of the qutrit density matrix $\mathbf{%
\rho }$. This definition of $D$ applies to pure and mixed states as well.
The behavior of $D$ allows us to verify that not only the populations but
also the phases of the probability amplitudes of the target state $%
\left\vert \Psi _{f}\right\rangle $ are produced by the QHR. Indeed, as time
progresses, $D$ approaches zero, which implies that $\left\vert \Psi
(t)\right\rangle $ aligns with $\left\vert \Psi _{f}\right\rangle $.

%
%
%
%
%
%
%
%
%
\begin{figure}[tbp]
\includegraphics[height=90mm]{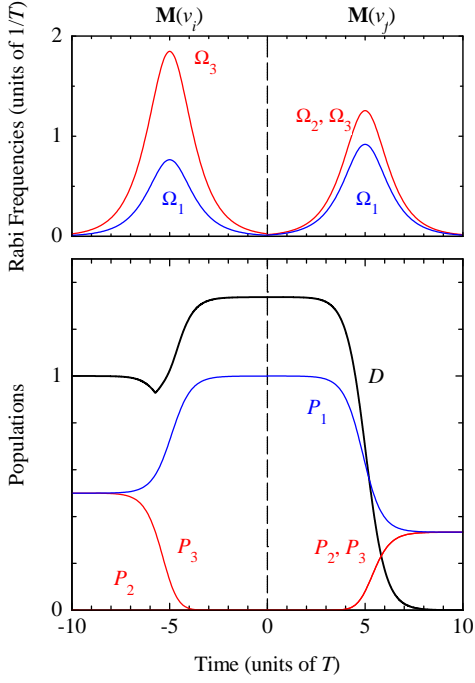}
\caption{(color online) Time evolution of the pulsed fields (top) and the
populations (bottom) for the transition (\protect\ref{2-3}) in a qutrit. We
have assumed sech pulse shapes and rms pulse area $A=2\protect\pi $ ($%
\protect\chi T=2$). The individual couplings $\protect\chi _{n}$ $(n=1,2,3)$
are given by the components of the standard QHRs (\protect\ref{2-3 standard}%
), each multiplied by $\protect\chi $. The detuning is $\Delta T=0$. The
thick curve shows the state mismatch (\protect\ref{mismatch}). }
\label{Fig 2-3 standard}
\end{figure}

\begin{figure}[tbp]
\includegraphics[height=90mm]{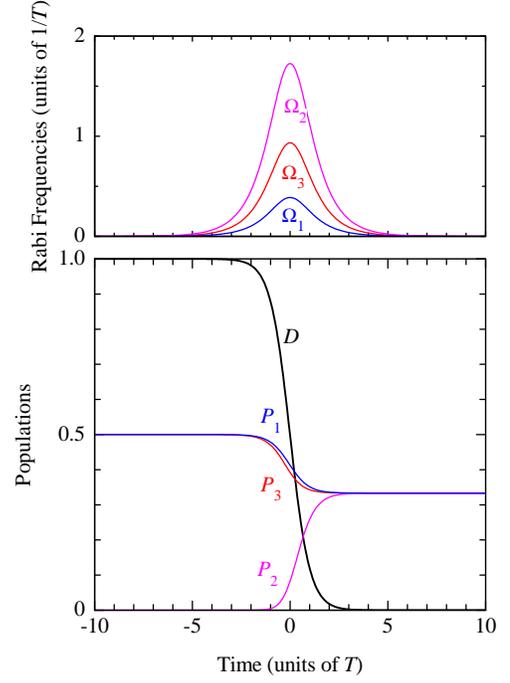}
\caption{(color online) Time evolution of the pulsed fields (top) and the
populations (bottom) for the transition (\protect\ref{2-3}) in a qutrit. We
have assumed sech pulse shapes and rms pulse area $A=2\protect\pi $ ($%
\protect\chi T=2$). The individual couplings $\protect\chi _{n}$ $(n=1,2,3)$
are given by the components of the generalized QHR (\protect\ref{2-3
generalized}), each multiplied by $\protect\chi $. The detuning is $\Delta T$
$=0.791$, which produces the desired phase $\protect\varphi =0.574\protect%
\pi $. The thick curve shows the state mismatch (\protect\ref{mismatch}).}
\label{Fig 2-3 generalized}
\end{figure}

In another example, we transfer a two-state superposition to a three-state
superposition,%
\begin{equation}
\frac{\left\vert 1\right\rangle +\left\vert 3\right\rangle }{\sqrt{2}}%
\longrightarrow \frac{\left\vert 1\right\rangle +e^{i\pi /3}\left\vert
2\right\rangle +e^{i\pi /7}\left\vert 3\right\rangle }{\sqrt{3}},
\label{2-3}
\end{equation}%
by two standard QHRs, $\mathbf{U}=\mathbf{M}(v_{f})\mathbf{M}(v_{i})$, with 
\begin{subequations}
\label{2-3 standard}
\begin{eqnarray}
\left\vert v_{i}\right\rangle  &=&\left[ -0.383,0,0.924\right] ^{T},
\label{1} \\
\left\vert v_{f}\right\rangle  &=&\left[ -0.460,0.628e^{i\pi
/3},0.628e^{i\pi /7}\right] ^{T},  \label{2}
\end{eqnarray}%
or by one generalized QHR, $\mathbf{U}=\mathbf{M}(v;\varphi )$, with 
\end{subequations}
\begin{equation}
\left\vert v\right\rangle =\left[ 0.194e^{0.213\pi i},0.863e^{-0.454\pi
i},0.467e^{-0.083\pi i}\right] ^{T},  \label{2-3 generalized}
\end{equation}%
and $\varphi =0.574\pi $. Figure \ref{Fig 2-3 standard} shows the time
evolution of the populations and the state mismatch (\ref{mismatch}) for a
standard-QHR implementation, and Fig. \ref{Fig 2-3 generalized} for
generalized QHR. In both cases, the mismatch $D$ vaniches, indicating the
creation of the desired superposition (\ref{2-3}). The generalized-QHR
implementation is clearly superior because it creates the target state in a
single step.

In conclusion of this section, we have demonstrated that any two pure
superposition qunit states can be connected by just a single generalized
QHR, or by two standard QHRs. This suggests that QHR, and particularly the
generalized version, is the most convenient and efficient tool for
pure-to-pure state navigation in Hilbert space. We now turn our attention to
the problem of connecting two arbitrary mixed states.

\section{Coherent navigation between mixed states\label{Sec-mixed}}

A mixed qunit state can be described by its density matrix $\mathbf{\rho }$,
whose spectral decomposition reads 
\begin{equation}
\mathbf{\rho }=\sum_{n=1}^{N}r_{n}\left\vert \psi _{n}\right\rangle
\left\langle \psi _{n}\right\vert .  \label{dm}
\end{equation}%
The eigenvalues $r_{n}$ of $\mathbf{\rho }$ satisfy $\sum_{n=1}^{N}r_{n}=1$,
and $\left\vert \psi _{n}\right\rangle $ are the orthonormalized $\left(
\langle \psi _{k}\left\vert \psi _{n}\right\rangle =\delta _{kn}\right) $
complex eigenvectors of $\mathbf{\rho }$. The density matrix is hermitian;
hence it can be parameterized by $N^{2}-1$ real parameters.

A hermitian Hamiltonian induces unitary evolution between an initial mixed
state $\mathbf{\rho }_{i}$ and a final state $\mathbf{\rho }_{f}$, 
\begin{equation}
\mathbf{\rho }_{f}=\mathbf{U}\mathbf{\rho }_{i}\mathbf{U}^{\dagger }.
\label{udmu1}
\end{equation}%
A unitary evolution does not change the eigenvalues $\left\{ r_{n}\right\}
_{n=1}^{N}$, which are therefore dynamic invariants, which is easily seen
from Eq. (\ref{udmu1}) (as an equivalent set of dynamic invariants one can
take the set $\left\{ \text{Tr}\mathbf{\rho }^{n}\right\} _{n=1}^{N}$).
Therefore, a unitary propagator $\mathbf{U}$ can only connect mixed states
with the same set of invariants $\left\{ r_{n}\right\} _{n=1}^{N}$. In order
to connect mixed states with different invariants we need an incoherent
process; we shall return to this problem in the next section. Here we shall
find the solution to the problem of linking two mixed states with the same
invariants.

Because the eigenvalues $\left\{ r_{n}\right\} _{n=1}^{N}$ of $\mathbf{\rho }%
_{i}$ and $\mathbf{\rho }_{f}$ are the same, we should have%
\begin{equation}
\mathbf{R}_{i}^{\dagger }\mathbf{\rho }_{i}\mathbf{R}_{i}=\mathbf{R}%
_{f}^{\dagger }\mathbf{\rho }_{f}\mathbf{R}_{f}=\mathbf{\rho }_{0},
\label{Rif}
\end{equation}%
where the unitary matrices $\mathbf{R}_{i}$ and $\mathbf{R}_{f}$ diagonalize
respectively $\mathbf{\rho }_{i}$ and $\mathbf{\rho }_{f}$, and $\mathbf{%
\rho }_{0}=$diag$\{r_{1},r_{2},\ldots ,r_{N}\}$. By replacing Eq. (\ref{Rif}%
) into Eq. (\ref{udmu1}) we find 
\begin{subequations}
\begin{eqnarray}
\mathbf{\rho }_{0} &=&\mathbf{D}\mathbf{\rho }_{0}\mathbf{D}^{\dagger },
\label{DrhoD} \\
\mathbf{D} &=&\mathbf{R}_{f}^{\dagger }\mathbf{UR}_{i}.  \label{D=RUR}
\end{eqnarray}%
Because $\mathbf{D}$ is a unitary matrix we find $\mathbf{\rho }_{0}\mathbf{D%
}=\mathbf{D}\mathbf{\rho }_{0}$. Since $\mathbf{\rho }_{0}$ is diagonal, $%
\mathbf{D}$ must be diagonal too. There are no other restrictions on $%
\mathbf{D}$; hence $\mathbf{D}$ is an arbitrary diagonal matrix. It follows
from Eq. (\ref{D=RUR}) that the solution for the unitary propagator in Eq. (%
\ref{udmu1}) is given by 
\end{subequations}
\begin{equation}
\mathbf{U}=\mathbf{R}_{f}\mathbf{DR}_{i}^{\dagger }.  \label{U}
\end{equation}

The propagator (\ref{U}) is not unique; it depends on the choice of the
diagonal matrix $\mathbf{D}$. In particular, we can always choose $\mathbf{D}%
=\mathbf{I}$\textbf{.} Hence, the transfer between two mixed states requires
a general U($N$) propagator. The latter can be expressed as a product of $N-1
$ standard QHRs $\mathbf{M}(v_{n})$ ($n=1,2,\ldots ,N-1$) and a phase gate $%
\mathbf{\Phi }\left( \phi _{1},\phi _{2},\ldots ,\phi _{N}\right) $, Eq. (%
\ref{standard}), or by $N$ generalized QHRs $\mathbf{M}(v_{n};\varphi _{n})$
($n=1,2,\ldots ,N$), Eq. (\ref{generalized}) \cite{Ivanov}.

We take as an example a \emph{qutrit}, with the arbitrarily chosen initial
and final density matrices 
\begin{subequations}
\label{rho-mixed}
\begin{equation}
\mathbf{\rho }_{i}=\left[ 
\begin{array}{ccc}
0.490 & 0.115e^{-0.789\pi i} & 0.158e^{0.107\pi i} \\ 
0.115e^{0.789\pi i} & 0.336 & 0.018e^{-0.675\pi i} \\ 
0.158e^{-0.107\pi i} & 0.018e^{0.675\pi i} & 0.175%
\end{array}%
\right] ,  \label{rho initial}
\end{equation}%
\begin{equation}
\mathbf{\rho }_{f}=\left[ 
\begin{array}{ccc}
0.298 & 0.022e^{0.689\pi i} & 0.033e^{0.319\pi i} \\ 
0.022e^{-0.689\pi i} & 0.180 & 0.177e^{0.909\pi i} \\ 
0.033e^{-0.319\pi i} & 0.177e^{-0.909\pi i} & 0.523%
\end{array}%
\right] .  \label{rho final}
\end{equation}%
These density matrices can be connected by the unitary propagator (\ref{U})
with $\mathbf{D}=\mathbf{I}$: $\mathbf{U}=\mathbf{R}_{f}\mathbf{R}%
_{i}^{\dagger }$. The latter can be expressed as a product of two standard
QHRs and one phase gate $\mathbf{U}=\mathbf{M}(v_{1})\mathbf{M}(v_{2})%
\mathbf{\Phi }$, with 
\end{subequations}
\begin{subequations}
\begin{eqnarray}
\left\vert v_{1}\right\rangle  &=&\left[ 0.612e^{0.532\pi
i},0.091e^{0.211\pi i},0.785e^{0.690\pi i}\right] ^{T},  \label{QHR mixed} \\
\left\vert v_{2}\right\rangle  &=&\left[ 0,0.533e^{-0.181\pi
i},0.846e^{0.859\pi i}\right] ^{T}, \\
\mathbf{\Phi } &=&\text{diag}\{e^{-0.468\pi i},e^{0.819\pi i},e^{-0.350\pi
i}\},
\end{eqnarray}%
or by three generalized QHRs, $\mathbf{U}=\mathbf{M}(v_{1};\varphi _{1})%
\mathbf{M}(v_{2};\varphi _{2})\mathbf{M}(v_{3};\varphi _{3})$, with 
\end{subequations}
\begin{subequations}
\label{mixed generalized}
\begin{eqnarray}
\left\vert v_{1}\right\rangle  &=&\left[ 0.721e^{0.659\pi
i},0.080e^{-0.209\pi i},0.689e^{0.270\pi i}\right] ^{T},  \label{mixed v1} \\
\left\vert v_{2}\right\rangle  &=&\left[ 0,0.813e^{0.469\pi
i},0.582e^{-0.261\pi i}\right] ^{T},  \label{mixed v2} \\
\left\vert v_{3}\right\rangle  &=&\left[ 0,0,1\right] ^{T},  \label{mixed v3}
\\
\varphi _{1} &=&-0.841\pi ,\text{ }\varphi _{2}=0.969\pi ,\text{ }\varphi
_{3}=-0.128\pi .  \label{mixed phi}
\end{eqnarray}

\begin{figure}[tbp]
\includegraphics[height=90mm]{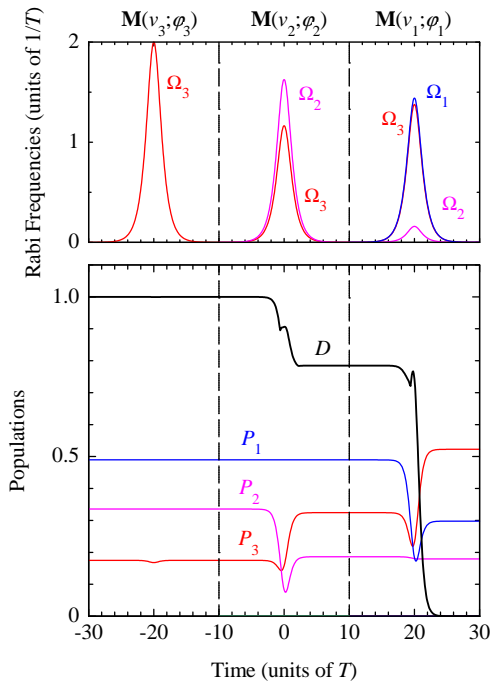}
\caption{(color online) Time evolution of the pulsed fields (top), the
populations and the state mismatch (bottom) for the transition between
states (\protect\ref{rho initial}) and (\protect\ref{rho final}) in a
qutrit. We have assumed sech pulse shapes and rms pulse area $A=2\protect\pi 
$ ($\protect\chi T=2$). The individual couplings $\protect\chi _{n}$ $%
(n=1,2,3)$ are given by the components of the generalized QHR (\protect\ref%
{mixed generalized}), each multiplied by $\protect\chi $. The detunings are $%
\Delta _{1}T=-0.255$, $\Delta _{2}T=0.049$, and $\Delta _{3}T=-4.918$, which
produce the QHR phases (\protect\ref{mixed phi}).}
\label{Fig-mixed}
\end{figure}

Figure \ref{Fig-mixed} shows the respective time evolution of the
populations and the state mismatch (\ref{mismatch}) for the generalized-QHR
realization (\ref{mixed generalized}). The first QHR $\mathbf{M}%
(v_{3};\varphi _{3})$ does not cause population changes because it is in
fact a phase gate. As time progresses, the mismatch decreases and the target
density matrix (\ref{rho final}) is approached.

\section{Synthesis of arbitrary preselected mixed states\label%
{Sec-engineering}}

As it was shown in the previous sections, by applying one or more QHRs one
can connect any two arbitrary pure states, or two arbitrary mixed states
with the same dynamic invariants $\left\{ r_{n}\right\} _{n=1}^{N}$. Mixed
states with different invariants cannot be connected by coherent hermitian
evolution because these invarants are constants of motion. Hence in order to
connect mixed states with different invariants we need a mechanism with
non-hermitian dynamics, which can alter the dynamic invariants.

In this section we shall describe two techniques for engineering an
arbitrary mixed state, starting from a single pure state. This is the most
interesting special case of the general problem of connecting two arbitrary
mixed states, because the initial state can be prepared routinely by optical
pumping. Moreover, the general mixed-to-mixed problem can be reduced to the
single-to-mixed problem by optically pumping the initial mixed state into a
single state.

The two techniques use a combination of coherent and incoherent evolutions.
The coherent evolution uses QHRs, whereas the incoherent non-hermitian
evolution is induced either by pure dephasing or spontaneous emission. We
shall consider the two techniques separately.

\subsection{Using dephasing}

We assume that the qunit is initially in the single qunit state $\mathbf{%
\rho }_{i}=\left\vert i\right\rangle \left\langle i\right\vert $, and we
wish to transform the system to an arbitrary mixed state $\mathbf{\rho }_{f}$%
. Let us denote the eigenvalues of $\mathbf{\rho }_{f}$ by $r_{n}$ ($%
n=1,2,\ldots ,N$). We proceed as follows.

\begin{itemize}
\item First, using the prescription from Sec. \ref{Sec-pure}, we apply a
single QHR to transfer state $\left\vert i\right\rangle $ to a pure
superposition state, in which the populations are equal to the eigenvalues
of $\mathbf{\rho }_{f}$: $\rho _{nn}=r_{n}$ ($n=1,2,\ldots ,N$). The phases
of this superposition are irrelevant.

\item In the second step we switch the dephasing on and let all coherences
decay to zero. This can be done, for example, by using phase-fluctuating
far-off-resonance laser fields. In the end of this process, the density
matrix will be diagonal, with the eigenvalues $r_{n}$ of $\mathbf{\rho }_{f}$
on the diagonal, which implies that it will have the same dynamic invariants
as $\mathbf{\rho }_{f}$.

\item The third step is to connect this intermediate state to the desired
state $\rho _{f}$ by a sequence of QHRs, as explained in the previous Sec. %
\ref{Sec-mixed}.
\end{itemize}

In summary, we need \emph{three} steps: a single QHR, a dephasing process,
and a sequence of $N$ QHRs. Figure \ref{Fig-engineering} shows the evolution
of the populations and the state mismatch (\ref{mismatch}) during the
engineering of the mixed state (\ref{rho final}) by the dephasing technique.
The first step is the single QHR $\mathbf{M}(v)$, with QHR vector
\end{subequations}
\begin{equation}
\left\vert v\right\rangle =\left[ -0.336,0.816,0.471\right] ^{T},
\label{engineering v}
\end{equation}%
which transfers the single initial state $\left\vert 1\right\rangle $ to the
pure superposition state 
\begin{equation}
\mathbf{\rho }_{1}=\left[ 
\begin{array}{ccc}
0.6 & \sqrt{0.18} & \sqrt{0.06} \\ 
\sqrt{0.18} & 0.3 & \sqrt{0.03} \\ 
\sqrt{0.06} & \sqrt{0.03} & 0.1%
\end{array}%
\right] .
\end{equation}%
The second step is the pure dephasing process, which nullifies all
coherences and leaves the density matrix in a diagonal form,%
\begin{equation}
\mathbf{\rho }_{2}=\text{diag}\left\{ 0.6,0.3,0.1\right\} .
\end{equation}%
The third step is a sequence of two generalized QHRs, which transfer $%
\mathbf{\rho }_{2}$ into the desired final density matrix $\mathbf{\rho }_{f}
$, Eq. (\ref{rho final}). The QHR components read 
\begin{subequations}
\label{mixed engineering}
\begin{eqnarray}
\left\vert v_{1}\right\rangle  &=&\left[ 0.689e^{0.454\pi
i},0.280e^{0.436\pi i},0.668e^{-0.477\pi i}\right] ^{T},
\label{engineering v1} \\
\left\vert v_{2}\right\rangle  &=&\left[ 0,0.793e^{0.740\pi
i},0.609e^{0.025\pi i}\right] ^{T},  \label{engineering v2} \\
\varphi _{1} &=&0.950\pi ,\quad \varphi _{2}=-0.760\pi .
\label{engineering phases}
\end{eqnarray}

\begin{figure}[tbp]
\includegraphics[width=70mm]{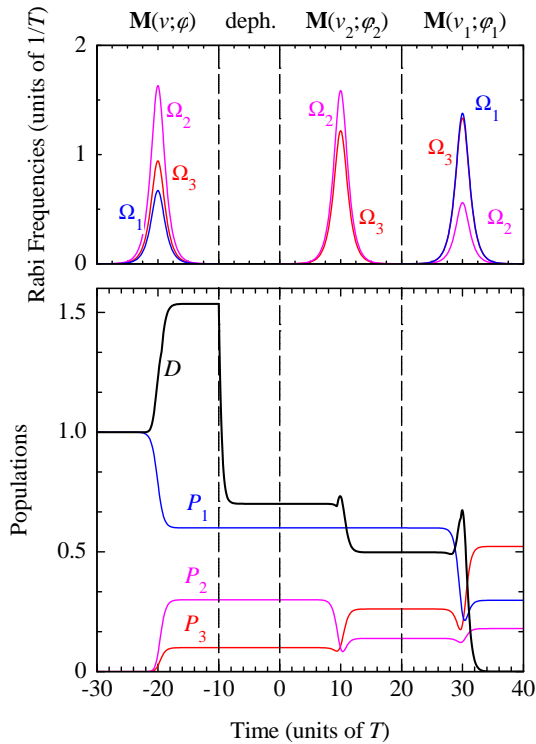}
\caption{(color online) Time evolution of the pulsed fields (top), the
populations and the state mismatch (\protect\ref{mismatch}) (bottom) for
mixed state engineering in a qutrit. The qutrit starts in state $\left\vert
1\right\rangle $ and the target final state is given by Eq. (\protect\ref%
{rho final}). We have assumed sech pulse shapes and rms pulse area $A=2%
\protect\pi $ ($\protect\chi T=2$). The individual couplings $\protect\chi %
_{n}$ $(n=1,2,3)$ are given by the components of the generalized QHR (%
\protect\ref{mixed engineering}), each multiplied by $\protect\chi $. The
detunings are $\Delta _{1}T=0.072$ and $\Delta _{2}T=-0.396$, which produce
the desired QHR phases (\protect\ref{engineering phases}). The dephasing
rate is $\Gamma =2/T$.}
\label{Fig-engineering}
\end{figure}

\subsection{Using spontaneous emission}

In the method, which uses spontaneous emission, we start again in a single
qunit state $\mathbf{\rho }_{i}=\left\vert i\right\rangle \left\langle
i\right\vert $, and the target is the arbitrary mixed state $\mathbf{\rho }%
_{f}$. The procedure now consists of only two steps: incoherent and
coherent. It is particularly well suited for a qutrit, which we shall
describe, although it is readily extended to more states. This method
requires a closed qunit-ancilla transition; if the ancilla state can decay
to other levels then the fidelity will be reduced accordingly.

It is possible here to apply directly the incoherent step, which produces a
density matrix with the desired final dynamic invariants, without the need
to prepare first a coherent qunit superposition, as in the dephasing method
above. The idea is to use laser-induced spontaneous emission from the
ancilla excited state to prepare a completely incoherent superposition of
the qunit states with populations $\rho _{nn}$ equal to the eigenvalues $%
r_{n}$ of $\mathbf{\rho }_{f}$, 
\end{subequations}
\begin{equation}
\mathbf{\rho }=\sum_{n=1}^{3}r_{n}\left\vert n\right\rangle \left\langle
n\right\vert .  \label{rho intermediate}
\end{equation}%
For this we apply a sequence of appropriately chosen laser pulses from the
qunit states to the excited state, which decays back to the qunit states and
redistridutes the population among them.

There are various scenarios possible, which can produce the desired
incoherent qunit superposition. Here we describe a scenario which looks
particularly simple and easy to implement for the qutrit formed of the
magnetic sublevels $M=-1,0,1$ of a $J=1$ level and an ancilla excited level
with $J=0$ (this implies also equal spontaneous decay branch ratios from the 
$J=0$ level to the $M$ sublevels of the qutrit). For definiteness, and
without loss of generality, we assume that the eigenvalues of $\mathbf{\rho }%
_{f}$ are ordered as $r_{1}\geqq r_{2}\geqq r_{3}$. We need three pulses: a
short pulse from state $\left\vert 1\right\rangle $, a long pulse from state 
$\left\vert 3\right\rangle $ and again a short pulse from state $\left\vert
1\right\rangle $ (here short and long are related to the lifetime of the
excited state).

The short pulse from the initially populated state $\left\vert
1\right\rangle $, with excitation probability $p_{1}$, transfers population $%
p_{1}$ to the excited state, 1/3 of which decays back to each of the qutrit
states. The ensuing density matrix reads 
\begin{equation}
\mathbf{\rho }_{1}=\text{diag}\left\{ 1-\frac{2}{3}p_{1},\frac{1}{3}p_{1},%
\frac{1}{3}p_{1}\right\} .  \label{rho1}
\end{equation}%
We then apply a sufficiently long pulse from state $\left\vert
3\right\rangle $, so that its population is completely depleted and
distributed among states $\left\vert 1\right\rangle $ and $\left\vert
2\right\rangle $. The resulting density matrix is 
\begin{equation}
\mathbf{\rho }_{2}=\text{diag}\left\{ 1-\frac{1}{2}p_{1},\frac{1}{2}%
p_{1},0\right\} .
\end{equation}%
We now apply again a short pulse from state $\left\vert 1\right\rangle $,
with a different probability $p_{2}$, and then wait for spontaneous emission
from the excited state. The result is%
\begin{eqnarray}
\mathbf{\rho }_{3} &=&\text{diag}\left\{ \left( 1-\frac{1}{2}p_{1}\right)
\left( 1-\frac{2}{3}p_{2}\right) ,\right.   \notag \\
&&\left. \frac{1}{2}p_{1}+\frac{1}{3}p_{2}\left( 1-\frac{1}{2}p_{1}\right) ,%
\frac{1}{3}p_{2}\left( 1-\frac{1}{2}p_{1}\right) \right\} .
\end{eqnarray}%
It is easy to show that in order to create the mixed state (\ref{rho
intermediate}) we should have the probabilities 
\begin{subequations}
\begin{eqnarray}
p_{1} &=&2\left( r_{2}-r_{3}\right) , \\
p_{2} &=&\frac{3r_{3}}{r_{1}+2r_{3}}.
\end{eqnarray}%
Because we assumed that $r_{1}\geqq r_{2}\geqq r_{3}$ the probabilities $%
p_{1}$ and $p_{2}$ belong to the interval $\left[ 0,1\right] $ and are
therefore well defined. Such probabilities can be produced by resonant
pulses with appropriate pulse areas $A_{n}$. These pulses should be short
compared to the lifetime of the excited state in order to avoid spontaneous
emission during their action. 

Once we have prepared the mixed qutrit state (\ref{rho intermediate}), which
has the same invariants as $\mathbf{\rho }_{f}$, we can apply QHRs to
transfer this state into the desired final state $\mathbf{\rho }_{f}$, as
described in Sec. \ref{Sec-mixed}.

\section{Conclusions\label{Sec-conclusions}}

In this paper we have proposed a technique, which allows to connect any two
quantum superposition states, pure or mixed, of an $N$-state atom. This
solution of the inverse problem in quantum mechanics contains two stages:
(i) mathematical derivation of the propagator that links the desired initial
and final density matrices, and (ii) physical realization of this
propagator. In the most general case of arbitrary mixed states, the
implementations combine coherent hermitian and incoherent non-hermitian
interactions induced by pulsed laser fields. In general, the propagator is
not unique, which reflects the multitude of paths between two qunit states;
this also allows for some flexibility in the choice of most convenient path.

The physical realization uses an $N$-pod configuration of $N$ lower states,
forming the qunit, and an ancillary upper state. It is particularly
convenient for a qutrit, where the $N=3$ states are the magnetic sublevels
of a $J=1$ level and the ancilla state is a $J=0$ level. Then only a single
tunable laser is needed to provide the necessary polarized laser pulses.

The hermitian part uses a sequence of sets of short coherent laser pulses
with appropriate pulse areas and detunings. For each set, the propagator of
the $N$-pod represents a quantum Householder reflection (QHR). A sequence of 
\emph{at most} $N$ suitably chosen QHRs can synthesize any desired unitary
propagator.

We have shown that two arbitrary preselected \emph{pure} superposition
states can be connected by a \emph{single} QHR only, because the respective
propagator has exactly the QHR symmetry. Two mixed states, with the same set
of dynamic invariants, require a general U($N$) transformation, which can be
realized by at most $N$ QHRs. This is a significant improvement over the
existing setups involving $O(N^{2})$\ operations, which can be crucial in
making quantum state engineering and operations with qunits experimentally
feasible.

The most general case of two arbitrary mixed states with different dynamic
invariants requires an incoherent step, which equalizes the invariants of
the initial density matrix to those of the final density matrix. We have
demonstrated how this can be done by using pure dephasing or spontaneous
decay of the ancillary upper state. Once the invariants are equalized, the
problem is reduced to the one of connecting two mixed states with the same
invariants, which, as explained above, can be done by at most $N$ QHRs. This
method has been described for a qutrit, but it is easily generalized to an
arbitrary qunit.

The present results can have important applications in the storage of
quantum information. For example, a qubit can encode two continuous
parameters: the population ratio of the two qubit states and the relative
phase of their amplitudes.\ A qunit in a \emph{pure }state can encode $2(N-1)
$\ parameters ($N-1$\ populations and $N-1$\ relative phases), i.e. by using
qunits information can be encoded in significantly fewer particles than with
qubits. Moreover, a \emph{mixed }qunit state can encode as many as $N^{2}-1$
real parameters. This may be particularly interesting if the number of
particles that can be used is restricted, e.g., due to decoherence \cite{QI}.

\acknowledgments

This work is supported by the European Union's ToK project CAMEL and RTN
project EMALI, and the Alexander von Humboldt Foundation.

\end{subequations}


\begin{thebibliography}{99}
\bibitem{Shore} B.W. Shore, \emph{The Theory of Coherent Atomic Excitation}
(Wiley, New York, 1990).

\bibitem{ARPC} N.V. Vitanov, T. Halfmann, B.W. Shore, and K. Bergmann, Ann.
Rev. Phys. Chem. \textbf{52}, 763 (2001).

\bibitem{STIRAP} K. Bergmann, H. Theuer and B.W. Shore, Rev. Mod. Phys. 
\textbf{70}, 1003 (1998); N.V. Vitanov, M. Fleischhauer, B.W. Shore and K.
Bergmann, Adv. At. Mol. Opt. Phys. \textbf{46}, 55 (2001).

\bibitem{QI} C.P. Williams and S.H. Clearwater, \emph{Explorations in
Quantum Computing} (Springer, Berlin, 1997); M.A. Nielsen and I.L. Chuang, 
\emph{Quantum Computation and Quantum Information} (Cambridge University
Press, Cambridge, 2000); D. Bouwmeester, A. Ekert, A. Zeilinger, \emph{The
Physics of Quantum Information: Quantum Cryptography, Quantum Teleportation,
Quantum Computation} (Springer, Berlin, 2001).

\bibitem{Karpati} A. Karpati, Z. Kis, and P. Adam, Phys. Rev. Lett. \textbf{%
93}, 193003 (2004).

\bibitem{Kyoseva} E.S. Kyoseva and N.V. Vitanov, Phys. Rev. A \textbf{73},
023420 (2006).

\bibitem{Ivanov} P.A. Ivanov, E.S. Kyoseva and N.V. Vitanov, Phys. Rev A 
\textbf{74}, 022323 (2006).

\bibitem{Householder} A. S. Householder, J. ACM \textbf{5}, 339 (1958).

\bibitem{Householder applications} J.H. Wilkinson, Comput. J. \textbf{3}, 23
(1960); Numer. Math. \textbf{4}, 354 (1962); J.M. Ortega, Numer. Math. 
\textbf{5}, 211 (1963); D.J. Mueller, Numer. Math. \textbf{8}, 72 (1966).

\bibitem{qudits-SU(2)} M. Reck, A. Zeilinger, H.J. Bernstein, and P.
Bertani, Phys. Rev. Lett. \textbf{73}, 58 (1994). A. Muthukrishnan and C. R.
Stroud Jr., Phys. Rev. A \textbf{62}, 052309 (2000); G.K. Brennen, D.P.
O'Leary and S.S. Bullock, Phys. Rev. A \textbf{71}, 052318 (2005); S.S.
Bullock, D.P. O'Leary and G.K. Brennen, Phys. Rev. Lett. \textbf{94}, 230502
(2005); A.B. Klimov, R. Guzm\'{a}n, J.C. Retamal, and C. Saavedra, Phys.
Rev. A \textbf{67}, 062313 (2003).

\bibitem{RZ} N. Rosen and C. Zener, Phys. Rev. \textbf{40}, 502 (1932).

\bibitem{femto} T. Brixner, T. Pfeifer, G. Gerber, M. Wollenhaupt, and T.
Baumert, in \emph{Femtosecond Laser Spectroscopy}, edited by P. Hannaford
Springer, New York, 2005, Chap. 9.
\end{thebibliography}
\end{document}